\documentstyle[leqno,twoside,11pt]{article}

\catcode`\@=11

\def\footnotesize{\@setsize\footnotesize{11pt}\ixpt\@ixpt
       \abovedisplayskip \z@
      \belowdisplayskip\z@
     \abovedisplayshortskip\abovedisplayskip
    \belowdisplayshortskip\belowdisplayshortskip
\def\@listi{\leftmargin\leftmargini \topsep 3pt plus 1pt minus 1pt
     \parsep 2pt plus 1pt minus 1pt
    \itemsep \parsep}}

\skip\footins 12pt plus 12pt

\def\footnoterule{\kern3\p@  \hrule width 3em\vspace{3pt}} 

\def\ps@plain{\let\@mkboth\@gobbletwo
     \def\@oddfoot{{\hfil\small\thepage\hfil}}%
     \def\@oddhead{}
      \def\@evenhead{}\def\@evenfoot{}}

\def\ps@headings{\let\@mkboth\markboth
        \def\@oddfoot{}\def\@evenfoot{}%
        \def\@evenhead{{\rm\thepage}\hspace*{2pc}{\sc\leftmark}\hfil}%
        \def\@oddhead{\hfil{\noindent\sc\rightmark}\hspace*{2pc}{\rm\thepage}}%

\def\ps@myheadings{\let\@mkboth\@gobbletwo
 \def\@oddfoot{}\def\@evenfoot{}%
 \def\@oddhead{\hfil{\sc\rightmark}\hspace*{2pc}{\normalsize\rm\thepage}}%
 \def\@evenhead{{\normalsize\rm\thepage}\hspace*{2pc}{\sc\leftmark}\hfil}%
}}

\def\abstract{\if@twocolumn
\section*{Abstract}
\else \small
\begin{center}
{\bf Abstract\vspace{-.5em}\vspace{3pt}}
\end{center}
\quotation
\fi}
\def\endabstract{\if@twocolumn\else\endquotation\fi}

\def\newproof#1{\@nprf{#1}}

\def\@nprf#1#2{\@xnprf{#1}{#2}}

\def\@xnprf#1#2{\expandafter\@ifdefinable\csname #1\endcsname
\global\@namedef{#1}{\@prf{#1}{#2}}\global\@namedef{end#1}{\@endproof}}

\def\@prf#1#2{\@xprf{#1}{#2}}

\def\@xprf#1#2{\@beginproof{#2}{\csname the#1\endcsname}\ignorespaces}

\def\newalgorithm#1{\@ifnextchar[{\@oalg{#1}}{\@nalg{#1}}}

\def\@nalg#1#2{%
\@ifnextchar[{\@xnalg{#1}{#2}}{\@ynalg{#1}{#2}}}

\def\@xnalg#1#2[#3]{\expandafter\@ifdefinable\csname #1\endcsname
{\@definecounter{#1}\@addtoreset{#1}{#3}%
\expandafter\xdef\csname the#1\endcsname{\expandafter\noexpand
  \csname the#3\endcsname \@thmcountersep \@thmcounter{#1}}%
\global\@namedef{#1}{\@alg{#1}{#2}}\global\@namedef{end#1}{\@endalgorithm}}}

\def\@ynalg#1#2{\expandafter\@ifdefinable\csname #1\endcsname
{\@definecounter{#1}%
\expandafter\xdef\csname the#1\endcsname{\@thmcounter{#1}}%
\global\@namedef{#1}{\@alg{#1}{#2}}\global\@namedef{end#1}{\@endalgorithm}}}

\def\@oalg#1[#2]#3{\expandafter\@ifdefinable\csname #1\endcsname
  {\global\@namedef{the#1}{\@nameuse{the#2}}%
\global\@namedef{#1}{\@alg{#2}{#3}}%
\global\@namedef{end#1}{\@endalgorithm}}}

\def\@alg#1#2{\refstepcounter
    {#1}\@ifnextchar[{\@yalg{#1}{#2}}{\@xalg{#1}{#2}}}

\def\@xalg#1#2{\@beginalgorithm{#2}{\csname the#1\endcsname}\ignorespaces}
\def\@yalg#1#2[#3]{\@opargbeginalgorithm{#2}{\csname
       the#1\endcsname}{#3}\ignorespaces}

\def\@beginproof#1{\rm {\it #1.\ }}
\def\@endproof{\outerparskip 0pt\endtrivlist}

\def\@begintheorem#1#2{\it {\sc #1\ #2.\ }}
\def\@opargbegintheorem#1#2#3{\it
      {\sc #1\ #2\ (#3).\ }}
\def\@endtheorem{\outerparskip 0pt\endtrivlist}

\def\@beginalgorithm#1#2{\rm \trivlist \item[\hskip \labelsep{\sc #1\ #2.}]}
\def\@opargbeginalgorithm#1#2#3{\rm \trivlist
      \item[\hskip \labelsep{\sc #1\ #2.\ (#3)}]}
\def\@endalgorithm{\outerparskip 6pt\endtrivlist}

\newskip\outerparskip

\def\trivlist{\parsep\outerparskip
  \@trivlist \labelwidth\z@ \leftmargin\z@
  \itemindent\parindent \def\makelabel##1{##1}}

\def\@trivlist{\topsep=0pt\@topsepadd\topsep
  \if@noskipsec \leavevmode \fi
  \ifvmode \advance\@topsepadd\partopsep \else \unskip\par\fi
  \if@inlabel \@noparitemtrue \@noparlisttrue
    \else \@noparlistfalse \@topsep\@topsepadd \fi
    \advance\@topsep \parskip
  \leftskip\z@\rightskip\@rightskip \parfillskip\@flushglue
  \@setpar{\if@newlist\else{\@@par}\fi}%
  \global\@newlisttrue \@outerparskip\parskip}

\def\endtrivlist{\if@newlist\@noitemerr\fi
   \if@inlabel\indent\fi
   \ifhmode\unskip \par\fi
   \if@noparlist \else
      \ifdim\lastskip >\z@ \@tempskipa\lastskip \vskip -\lastskip
         \advance\@tempskipa\parskip \advance\@tempskipa -\@outerparskip
         \vskip\@tempskipa
   \fi\@endparenv\fi
   \vskip\outerparskip}

 \newproof{@proof}{Proof}

 \newtheorem{@theorem}{Theorem}[section]

\newproof{Example}{Example}
\newproof{Method}{Method}
\newproof{Exercise}{Exercise}

 \def\@figtxt{figure}
\long\def\@makecaption#1#2{\small
\setlength{\parindent}{18pt}
\baselineskip 14pt
 \ifx\@captype\@figtxt
 \vskip 10pt
 \setbox\@tempboxa\hbox{{\sc #1} {\it #2}}
 \ifdim \wd\@tempboxa >\hsize {\sc #1} {\it #2}\par \else \hbox
to\hsize{\hfil\box\@tempboxa\hfil}%
 \fi\else\hbox to\hsize{\hfil{\sc #1}\hfil}%
 \setbox\@tempboxa\hbox{{\it #2}}%
 \ifdim \wd\@tempboxa >\hsize {\it #2}\par \else
 \hbox to \hsize{\hfil\box\@tempboxa\hfil}\fi
 \vskip 10pt
 \fi}

\def\fnum@figure{\par\sc Fig. \thefigure.\ }
\def\fnum@table{\small \sc Table \thetable}

\def\section{\@startsection {section}{1}{\z@}{-3.5ex plus -1ex minus
 -.2ex}
{2pt}{\large\bf}}
\def\subsection{\@startsection{subsection}{2}{\z@}{-3.25ex plus -1ex minus
 -.2ex}
{2pt}{\large\bf}}
\def\subsubsection{\@startsection {subsubsection}{3}{\z@}{1.3ex plus .5ex minus
    .2ex}{-.5em plus -.1em}{\normalsize\bf}}

\def\thebibliography#1{%
\parindent 0em
\vspace{9pt}
\begin{flushleft}\large\bf {References}\end{flushleft}
\addvspace{3pt}\nopagebreak\list
{[\arabic{enumi}]} {\settowidth\labelwidth{[#1]}
\leftmargin\labelwidth
\leftmargin=17pt
 \advance\leftmargin\labelsep
 \usecounter{enumi}\@bibsetup}
\def\newblock{\hskip .11em plus .33em minus -.07em}
 \sloppy\clubpenalty4000\widowpenalty4000
 \sfcode`\.=1000\relax}

\def\@bibsetup{
\itemsep=0pt \parsep=0pt
\small}

\def\theindex{\@restonecoltrue\if@twocolumn\@restonecolfalse\fi
\columnseprule \z@
\columnsep 35pt\twocolumn[\chapter*{Index}]
 \parskip\z@ plus .3pt\relax\let\item\@idxitem}

\def\printindex{\cleardoublepage\markboth{INDEX}{INDEX}
\addcontentsline{toc}{chapter}{Index}\@input{\jobname.ind}}

\ps@headings

\def\hg {{\cal HG}_{x_0}}
\def\agb {\overline {{\cal A}/{\cal G}}}
\def\ag {{\cal A}/{\cal G}}
\def\a{\alpha}
\def\b{\beta}
\def\d{\delta}

\def\Lapop{\displaystyle{{\hbox to 0pt{$\sqcup$\hss}}\sqcap}}

\def\R{{\rm I\!R}}

\def\Q{{\mathchoice
{\setbox0=\hbox{$\displaystyle\rm Q$}\hbox{\raise 0.15\ht0\hbox to0pt
{\kern0.4\wd0\vrule height0.8\ht0\hss}\box0}}
{\setbox0=\hbox{$\textstyle\rm Q$}\hbox{\raise 0.15\ht0\hbox to0pt
{\kern0.4\wd0\vrule height0.8\ht0\hss}\box0}}
{\setbox0=\hbox{$\scriptstyle\rm Q$}\hbox{\raise 0.15\ht0\hbox to0pt
{\kern0.4\wd0\vrule height0.7\ht0\hss}\box0}}
{\setbox0=\hbox{$\scriptscriptstyle\rm Q$}\hbox{\raise 0.15\ht0\hbox to0pt
{\kern0.4\wd0\vrule height0.7\ht0\hss}\box0}}}}
\def\C{{\mathchoice
{\setbox0=\hbox{$\displaystyle\rm C$}\hbox{\hbox to0pt
{\kern0.4\wd0\vrule height0.9\ht0\hss}\box0}}
{\setbox0=\hbox{$\textstyle\rm C$}\hbox{\hbox to0pt
{\kern0.4\wd0\vrule height0.9\ht0\hss}\box0}}
{\setbox0=\hbox{$\scriptstyle\rm C$}\hbox{\hbox to0pt
{\kern0.4\wd0\vrule height0.9\ht0\hss}\box0}}
{\setbox0=\hbox{$\scriptscriptstyle\rm C$}\hbox{\hbox to0pt
{\kern0.4\wd0\vrule height0.9\ht0\hss}\box0}}}}

\font\fivesans=cmss10 at 4.61pt
\font\sevensans=cmss10 at 6.81pt
\font\tensans=cmss10
\newfam\sansfam
\textfont\sansfam=\tensans\scriptfont\sansfam=\sevensans\scriptscriptfont
\sansfam=\fivesans
\def\sans{\fam\sansfam\tensans}
\def\Z{{\mathchoice
{\hbox{$\sans\textstyle Z\kern-0.4em Z$}}
{\hbox{$\sans\textstyle Z\kern-0.4em Z$}}
{\hbox{$\sans\scriptstyle Z\kern-0.3em Z$}}
{\hbox{$\sans\scriptscriptstyle Z\kern-0.2em Z$}}}}
\def\semi{\bigcirc\kern-1em{s}\;}

\newcount\foot
\def\note#1{\footnote{${}^{\number\foot}$}{\ftn #1}\advance\foot by 1}

\def\frac#1#2{{#1\over #2}}
\def\text#1{\quad{\hbox{#1}}\quad}

\font\ftn=cmr8 scaled\magstephalf

\font\it=cmti10 scaled\magstephalf
\font\bf=cmbx10 scaled\magstephalf

\def\YM{Yang-Mills \ }
\def\ha{\cal HA}
\def\hab{\overline{\ha}}

\begin{document}
\cleardoublepage
\pagestyle{myheadings}

\title{
Integration on the Space of Connections Modulo Gauge Transformations}
\author{Abhay Ashtekar\thanks{Center for Gravitational Physics and
Geometry, Physics Department, Penn State University, University Park,
PA 16802-6300, USA. Supported by NSF Grant PHY93-96246 and the Eberly
research fund of Penn State University.}
\and
Donald Marolf$^\ast$
\and
Jos\'e Mour\~ao\thanks{Center for Gravitational Physics and Geometry,
Penn State University, University Park, PA 16802-6300, USA.
Supported by funds provided by PSU and by a NATO grant 9/C/93/PO.}
\thanks{On leave of absence from Dept. F\'{\i}sica, Inst. Sup.
T\'{e}cnico, 1096 Lisboa, PORTUGAL. }}
\date{March 1994}
\maketitle
\markboth{Ashtekar and Marolf and Mour\~ao} {integration on the space
of connections modulo gauge transformations}

\pagenumbering{arabic}

\begin{abstract}
A summary of the known results on integration theory
on the space of connections modulo gauge transformations
is presented and its significance to quantum theories
of gauge fields and gravity is discussed. The emphasis is
on the underlying ideas rather than the technical subtleties.

\end{abstract}

\section{Introduction}

The purpose of this article is to provide theoretical physicists with a
summary  of some of the recent mathematical developments in the
integration theory on the space $\ag$ of connections modulo gauge
transformations. In a related article in this volume, Jerzy
Lewandowski has summarized the situation with differential calculus.
Together, these results provide a framework for a non-perturbative
quantization of gauge theories, particularly the ones in which
diffeomorphism invariance plays a significant role. The key idea is to
take the non-linear character of $\ag$ seriously by incorporating it
into the very fabric of the construction right in the first step.
This is a notable departure from perturbative as well as
the usual constructive
field theoretic approaches which are tailored to systems in which
configuration spaces are linear (see \cite{GJ,Riv}). The present
approach, on the other
hand, provides a continuum analog of lattice gauge theories which also
handle the ``kinematical non-linearities'' squarely.

Most of the material presented here has appeared (or is about to
appear) in the mathematical literature. Therefore, here, we will omit
detailed proofs and avoid the subtle technical issues. Rather, our
goal is to provide motivations behind various constructions, explain
the key ideas on which proofs are based, present the statements of
main results and supply heuristic arguments and pictures to make the
results and their ramifications intuitively plausible to the
theoretical physics community. In particular, we will emphasize the
points where caution is needed and explain in greater detail the
results that may seem surprising or counter intuitive at first.

Let us begin with the motivation.

Recall first that gauge potentials or connections feature prominently
in the current descriptions of all three fundamental interactions of
particle physics. Indeed, the principle of local gauge invariance
played a crucial role in the very unification of the electromagnetic
and weak interactions \cite{gi,JZ}. The same principle underlies
QCD, the best theory of strong interactions available today
\cite{JZ,QCD}.
General relativity, on the other hand, is generally regarded as
a dynamical theory of metrics and gravity therefore seemed to stand
apart from other basic interactions. However, in mid-eighties, it was
realized that general relativity can also be thought of as a dynamical
theory of connections \cite{AA87}. Not only does this formulation
bring gravity closer to other interactions but it also simplifies the
basic equations of the theory considerably. Over the last few years,
therefore, notable progress has occurred in non-perturbative quantum
gravity.  Furthermore, there has been a synergistic exchange of ideas
between quantum general relativity and quantum Yang-Mills theories
\cite{RS,BGP}. Finally, connections play an important role also in
topological field theories, e.g. Chern-Simons theories,
2+1-dimensional gravity, etc. [9-12], some of which are important for
the description of phase transitions in statistical physics and for
finding the continuum limit of the lattice formulation of field
theories. Thus, theories of connections play a dominant role in all
branches of fundamental physics.

Over the years, considerable progress has been made on quantization of
these theories. In particular, the standard model of particle physics
has been remarkably successful. Nonetheless, a number of important
problems remain, especially in the context of QCD. It is widely
believed that a satisfactory treatment of these problems will require
an understanding of non-perturbative phenomena. An outstanding example
of these are issues related to the physical properties of the true
vacuum state of the theory \cite{JZ,QCD}. Perhaps the most
significant non-perturbative approach to tackle such problems is
provided by lattice gauge theories \cite{JZ,QCD}. However the
difficulty in defining the continuum limit of these theories
indicates that new ideas and methods are still needed.  The
development of calculus on the space of connections modulo gauge
transformations opens up one such avenue.

Specifically, the theory of integration on $\ag$ will serve at least
four purposes. First, it is natural in the various physical theories
listed above to work in the connection representation where quantum
states arise as suitable functions $\Psi(A)$ of connections.  (In
theories where connections are configuration variables, one is led to
let $\Psi(A)$ be any complex-valued function, while in cases when
connections serve as phase space variables, $\Psi(A)$ are required to
be holomorphic.) These states are subject to the quantum Gauss
constraint which requires them to be gauge invariant:
\begin{equation}
\Psi(A) = \Psi(A^g)   \ ,
\end{equation}
where $A^g = g^{-1} A g + g^{-1}dg$. Thus, the physical states $\Psi$
project to functions on $\ag$. To endow them with the structure of a
Hilbert space, therefore, we need a measure $\mu$ on
$\ag$ that would allow us to compute the inner products:
\begin{equation}
\label{inp}
(\psi_1, \psi_2) = \int_{\ag} d\mu([A])\ \ \overline {\psi_1([A])}
\psi_2([A])
\end{equation}
The problem of finding a suitable measure has been resolved in lattice
gauge theories, 1+1-dimensional Yang-Mills theories (see e.g.
\cite{Baez1}), in 2+1-dimensional general relativity \cite{AAetal} and
certain topological field theories (such as the Chern-Simons theories
based on a group $IG$, the inhomogeneous version of a compact Lie
group $G$ \cite{AR2}). In all these cases, however, the space $\ag$ is
{\it finite} dimensional. For physically more interesting cases, such
as the Yang-Mills theory and general relativity in 3+1 dimensions,
$\ag$ is infinite dimensional and the problem is completely open.

The second motivation for developing integration theory comes from
certain problems which are peculiar to 3+1-dimensional general
relativity. In this case, to select the physical states of the theory,
one has to impose, in addition to the Gauss law, four additional
quantum constraints which ensure the diffeomorphism invariance of the
theory and whose expressions and actions are more complicated than that
of the Gauss law. Before imposing them, therefore, one faces the
problem of regularization of the formal expressions of the
corresponding operators. The availability of a ``kinematical Hilbert
space'' on which these operators can act would provide a natural setting
for this task. Having solved this problem, one needs a
criterion to select the suitably regular solutions which are to qualify
as physical states. The only natural procedure available today involves
the introduction of a rigged Hilbert space structure on the space of
functions on $\ag$ \cite{Haj}. Thus, integration theory on $\ag$ is needed
to resolve both problems.

The third motivation comes from the so called ``loop transform'' which
was first introduced by Gambini and Trias \cite{GT} in the context of
quantum Yang-Mills theory and later but independently by Rovelli and
Smolin \cite{RS} in the context of quantum gravity. The idea is to
mimic the Fourier transform which enables us, in quantum mechanics, to
pass from wave functions $\Psi(x)$ of position to the wave functions
$\psi(k)$ of momentum: \begin{equation} \psi(k) = \int_{\R} dx e^{ikx}
\Psi(x) \ , \end{equation} and construct a similar integral to pass
from the wave functions of connections to those of loops:
\begin{equation}
\label{lt} \psi (\gamma) = \int_{\ag} d\mu([A])\ \
T_\gamma([A]) \Psi([A]),
\end{equation}
where $T_\gamma$ denotes the
Wilson loop functional
\begin{equation}
\label{wlv}
T_\gamma ([A]) = {\rm Tr}\ {\cal P} exp \oint_\gamma A \cdot dx,
\end{equation}
and now plays the role of the integral kernel $e^{ikx}$ in the Fourier
transform. In the absence of a well-defined measure on $\ag$, of course,
the loop transform (\ref{lt}) is only a formal expression.
Nonetheless, as a heuristic device, it has already played a powerful
role in quantum \YM theories both in continuum \cite{GT,GT2} and on
lattice \cite{B,Lo} as well as in quantum general relativity
\cite{RS,BGP}.  In particular, in diffeomorphism invariant theories such as
general relativity, the resulting loop representation is easier to
handle than the original connection representation. Being
diffeomorphism invariant, the physical states become functionals of
knot classes of loops thereby bringing out a new role for knot theory
in physics. As we noted above, in situations in which $\ag$ is finite
dimensional, natural measures exist and these can be used to perform
the loop transform (\ref{lt}) and construct the loop representations
rigorously. In the physically more interesting cases when $\ag$ is
infinite dimensional, on the other hand, this avenue to the
construction of loop representations has not been available.

Finally, a proper integration theory is needed also in the Euclidean
approaches to quantum theories of connections.  If we let $\ag$ denote
the space of connections modulo gauge transformations on the Euclidean
space-time $M = \R \times \Sigma$ rather than on a Cauchy surface
$\Sigma$ (as was implicitly assumed so far) then measures on $\ag$
would provide a natural avenue to an Euclidean path integral
formulation which is well suited to handle the ``kinematical
non-linearities'' of the problem squarely.

With all these motivations at hand, it is natural to ask why the
integration theory on $\ag$ was not developed, say, twenty years ago.
What are the principal difficulties? The key problem is that in
general $\ag$ is a rather complicated, non-linear,
infinite-dimensional space and the difficulties arising from
non-linearities get coupled with those arising from the presence of
infinite dimensions. In the simple cases when only one of the
difficulties is present, the solution has ben available for some time
now. In 1+1-dimensional Yang-Mills theory or 2+1-dimensional general
relativity, for example, $\ag$ is non-linear but {\it finite}
dimensional and integration theory is straightforward. Similarly, in
the Abelian case (say the Maxwell theory), $\ag$ is
infinite-dimensional but {\it linear} and one can use a Gaussian
measure. It is not surprising that in the non-Abelian context, there
has been a temptation to try to ``iron out'' non-linearities with
gauge fixing and use methods that have been successful in the linear
context \cite{GJ,Riv}. To face the non-linearities head-on, on the
other hand, one needs to go beyond standard function spaces and
Gaussian measures and introduce new techniques to replace the familiar
ones.

The key idea in the approach we will discuss here is to achieve this
by exploiting the ``non-linear duality'' between connections and
loops. This program was initiated in '92 \cite{ai} and developed in a
series of papers over the last two years [21-24].
However, here, we will refrain from presenting the results in a
chronological order or from providing the various complementary
perspectives from which the constructions can be understood. Instead,
we shall present just one perspective which, we feel, is best suited
for the intended readership even though this will prevent us from doing
full justice to all the papers.

This paper is organized as follows. In Sect. 2 we review the
integration techniques used in field theories with a linear
configuration space. In Sect. 3 we develop analogous techniques for
the space $\ag$. In particular, using the non-linear duality between
connections and loops, we will obtain a non-linear extension of the
concepts of cylindrical measures and cylindrical functions. These are
the simplest functions we will be able to integrate. As a physically
relevant example, we will use the Haar measure on the group to define
a faithful, diffeomorphism-invariant measure on $\ag$
\cite{AL}. It turns out, however, that on $\ag$, this measure lacks a
technical property (called $\sigma$-additivity) with the result that
even though we begin with square-integrable functions on $\ag$, not
all elements of the completed Hilbert space can be represented as
functions on $\ag$.  In Sect. 4, we show that this situation can be
rectified by suitably extending $\ag$ to a space $\agb$ which contains
``distributional analogs of gauge-equivalent connections.'' Every
element of the quantum Hilbert space can be represented by a function
on $\agb$.  Thus, as is usual in quantum field theory, the domain
space of quantum states $\agb$ is significantly larger than the
classical configuration space $\ag$. We present an explicit
characterization of $\agb$: it can be obtained as a projective limit
of finite-dimensional spaces corresponding to gauge theories on finite
lattices \cite{mm}. This construction is rather general and can in
principle be carried out also in the case of a non-compact gauge group
$G$ (such as $SL(2, \C)$, the group relevant to gravity), effectively
{\it defining} $\agb$ in such cases.  In an Appendix, we present
another characterization of $\agb$ based on the theory of $C^\star$
algebras \cite{AL}; chronologically, it was this characterization that
first led to the integration theory on $\agb$ for $SU(N)$ gauge
fields.

\section{Linear case: Review}

We will first consider a scalar field in Minkowski space and then briefly
summarize the situation for the Maxwell field.

Consider scalar fields in $d+1$-dimensional Minkowski space satisfying the
equation \begin{equation}
\Lapop \phi - {\partial V(\phi) \over \partial \phi} = 0 \ ,
\end{equation}
where $V(\phi)$ is a self-interaction potential. It is natural
to choose, for the classical configuration space of the theory,
the linear space of all  suitably regular
field configurations (say $C^2$ and rapidly
decreasing  at
infinity)
at a given time:
\begin{equation}
{\cal C} = \{ \phi({\bf x}) \}   \ .
\end{equation}
Heuristically, one expects quantum states to be functions $\Psi(\phi)$
on ${\cal C}$ and the inner product to be given by an expression of the
type:
\begin{equation}
\label{sps}
(\Psi_1, \Psi_2) = \int_{\cal C} \overline{\Psi_1(\phi)} \Psi_2(\phi)
d\mu(\phi).
\end{equation}
Therefore, to construct the quantum theory, we need to introduce
suitable measures $\mu$ on ${\cal C}$ and develop integration theory.
\cite{GJ,Riv,GV}.

Let ${\cal L}$ denote the space of all test (or smearing) functions on
$\R^d$, i.e. the space of all continuous, infinitely differentiable
functions, which fall-off sufficiently rapidly at infinity. ${\cal L}$
is called the Schwarz space and its elements serve to label the
``configuration observables'': test functions $f$ in ${\cal L}$
define linear functionals on $\cal C$, $ F_f \ : \ {\cal C}\rightarrow
\R $, through
\begin{equation}
\label{sifu}
F_f(\phi) = < f , \phi > = \int_{\R^d} d^d{\bf x} f({\bf x}) \phi({\bf
x}) \ .
\end{equation}
These provide the simplest functions that one can introduce on $\cal
C$. The next simplest class is provided by the so called cylindrical
functions (see, e.g., \ \cite{GV,Ya}). These depend only on a finite
number of variables and can be defined as follows.  Let $V_n^*$ be a
finite dimensional subspace of ${\cal L}$ and $(e_1, \ldots , e_n)$ be
a basis on $V^*_n$.  (Here, the symbol $V$ stands for a linear space,
$n$, denotes its dimension, and the $*$ serves to remind us that the
elements of $V^*_n$ label configuration observables and are therefore
to be thought of as belonging to the dual of ${\cal C}$.)  Then we can
define certain projections $\pi_{e_1, \ldots , e_n} \ : \ {\cal C}
\rightarrow \R^n $ as follows:
\begin{equation}
\label{prol}
\pi_{e_1, \ldots , e_n}(\phi) =  \lbrace <e_1, \phi>, \ldots , <e_n,
\phi> \rbrace \ .
\end{equation}
A cylindrical function on ${\cal C}$ is then a function that can be
represented in the form
\begin{equation}
\label{cli}
f = F \circ \pi_{e_1, \ldots , e_n} \quad \hbox{or, equivalently}
\quad f(\phi) = F(<e_1, \phi>, \ldots , <e_n, \phi>),
\end{equation}
for some $\{e_1, \dots , e_n \}$ in $V_n^*$ and some function $F \ :
\ \R^n \rightarrow \C$. In other words $f$ is the pull-back
 to $\cal C$ of the
function $F $ on $\R^4$
(with respect to $\pi_{e_1, \ldots , e_n}$).
In this case, we will say that
$f$ is cylindrical with respect to $V_n^*$.  Clearly, the
representation (\ref{cli}) of $f$ is not unique. In particular, a
function which is cylindrical with respect to $V^*_n$ is cylindrical
with respect to any space $\widetilde V^*_m$ that contains $V^*_n$.

Cylindrical measures on $\cal C$ are, as the name suggests, measures
that enable one to integrate cylindrical functions (see
\cite{GV,Ya}).  As is natural from (\ref{cli}) a cylindrical measure $\mu$ on
$\cal C$ is in one-to-one correspondence with an infinite {\it
consistent} family $\{ \mu_{e_1, \ldots , e_n} \}$ of measures on the
finite dimensional spaces $\R^n$ (one for every finite, ordered set of
linearly independent vectors $\{ e_1, \ldots , e_n \}$):
\begin{equation}
\mu \leftrightarrow  \{ \mu_{e_1, \ldots , e_n} \} \ .
\end{equation}
Given such a family of measures, $\{ \mu_{e_1, \ldots , e_n} \}$,
a cylindrical function $f = F \circ \pi_{e_1, \ldots , e_n}$ on $\cal
C$ is said to be integrable with respect to $\mu$ if and only if $F$ is
integrable with respect to $\mu_{e_1, \ldots , e_n}$ and in this case
the integral of $f$ with respect to $\mu$ is defined to be
\begin{equation}
\label{cyme}
\int_{\cal C} d\mu(\phi) f (\phi) = \int_{\R^n} F(\eta_1, \ldots , \eta_n)
d\mu_{e_1, \ldots , e_n}(\eta_1, \ldots , \eta_n)    \ .
\end{equation}
Considering (\ref{cyme}) for every integrable cylindrical function
(including characteristic functions of sets) we effectively define the
the measure $\mu$. Next, let us examine the consistency conditions on
the permissible family $\{ \mu_{e_1, \ldots , e_n} \}$. These follow
directly from (\ref{cyme}). Consider first the simplest case when the
linear spaces $V^*_n$ and $\widetilde V_m^*$ with basis $\{ e_1,
\ldots , e_n\}$ and $\{ \tilde e_1, \ldots , \tilde e_m \}$ are such
that
       $$ V^*_n \cap \widetilde V^*_m = \{ 0 \}. $$
The consistency condition is that functions which are cylindrical with
respect to both $V^*_n$ and $\widetilde V^*_m$, should have the same
integral. Since such functions are necessarily constants, we obtain
\begin{equation}
\label{norm}
\int_{\cal C} d\mu \cdot 1 = \int_{\R^n}
d\mu_{e_1, \ldots , e_n} \cdot 1   =
\int_{\R^m}
d\mu_{\tilde e_1, \ldots , \tilde e_m}\cdot 1
\end{equation}
or
\begin{equation}
\mu({\cal C}) = \mu_{e_1, \ldots ,  e_n}(\R^n) =
\mu_{\tilde e_1, \ldots , \tilde e_m} (\R^m)   \ .
\end{equation}
Apart from these normalization conditions, the measures $\mu_{ e_1,
\ldots , e_n}$ and $\mu_{\tilde{e}_1, \ldots , \tilde{e}_n}$ can be
chosen arbitrarily. If however
$$
V^*_n( e_1 , \ldots ,  e_n)
 \cap \widetilde V^*_m (\tilde e_1 , \ldots , \tilde e_m) =
\hat V^*_k (\hat e_1 , \ldots , \hat e_k)
$$
with $k > 0$, then the requirement that the integral of functions $f$
which are cylindrical with respect to both $V^*_n$ and $\widetilde
V^*_m$, be well defined leads to further consistency conditions.
It is easy to see that these are satisfied if for every
$$
V^*_n( e_1 , \ldots ,  e_n) \subset \widetilde V^*_m (\tilde e_1 ,
\ldots , \tilde e_m)  \quad {\rm with}\quad
e_i = \sum_{j=1}^m L_{ij} \tilde e_j   \  ; \quad i = 1 , \ldots , n \ ;
 \quad m \geq n
$$
and for every function $f$ which is cylindrical with respect
to  $V^*_n( e_1 , \ldots ,  e_n)$, i.e. which satisfies
\begin{eqnarray}
f(\phi) &=& F(<e_1, \phi>, \ldots , <e_n, \phi>)
= F(<L_{1j} \tilde e_j, \phi>, \ldots , <L_{nj} \tilde e_j, \phi>)
\nonumber \\
 &=& \widetilde F(<\tilde e_1, \phi>, \ldots , <\tilde e_m, \phi>)
\end{eqnarray}
we have
\begin{equation}
\label{cons}
\int_{\R^n} F(\eta_1, \ldots , \eta_n) d\mu_{e_1, \ldots , e_n}(\eta_1,
\ldots , \eta_n)    = \int_{\R^m}
\widetilde F(\tilde \eta_1, \ldots , \tilde \eta_m)
d\mu_{\tilde e_1, \ldots , \tilde e_m}(\tilde \eta_1,
\ldots , \tilde \eta_m)   \ .
\end{equation}
Thus, every family of measures $\{\mu_{e_1, \ldots , e_n}\}$ on finite
dimensional spaces which satisfies the consistency conditions
(\ref{cons}) for every integrable function $F$, defines a cylindrical
measure on ${\cal C}$ via (\ref{cyme}).  Conversely, every cylindrical
measure $\mu$ on $\cal C$ defines, again through (\ref{cyme}), a
consistent family of measures $\{ \mu_{e_1, \ldots , e_n} \}$.

A particularly simple class of solutions to the consistency conditions
(\ref{cons}) is provided by the normalized gaussian measures on
$\R^n$.  The cylindrical measure defined with the help of a family of
gaussian measures on finite dimensional spaces is also called a
gaussian measure. (This is normalized so that $\mu({\cal C}) =1$.)
The text-book quantum theory of free fields is based on these
measures.  In particular, the Fock space of a free massive scalar
field can be obtained by Cauchy completion of the space of cylindrical
functions on ${\cal C}$ which are square-integrable with respect to
the gaussian measure with covariance given by ${1 \over 2} (- \Delta +
m^2)^{-{1 \over 2}}$.  One expects the interacting quantum scalar
field theories to arise from non-gaussian cylindrical measures
\cite{GJ}.  This expectation is borne out in 1+1 and 2+1 dimensional
Minkowski spaces.

Aside from the steps involved in defining cylindrical measures, at
first sight the final picture here seems remarkably similar to that
in non-relativistic quantum mechanics where the Hilbert space of
physical states is also obtained by taking the Cauchy completion of
suitably regular, square integrable functions on the classical
configuration space. There is, however, a key difference. In
non-relativistic quantum mechanics, every physical state ---i.e.,
every element of the completion--- can be represented as a function of
the classical configuration space. This is {\it not} the case in field
theory. Now, not all elements of the completion can be regarded as
functions on ${\cal C}$.  They {\it can}, however, be represented as
functions on a larger space, that of the {\it distributions} on the
constant time surface. Thus, the domain space of quantum states is no
longer the classical configuration space; it is larger.

The overall situation is similar in the case of Maxwell fields in
Minkowski space. The classical configuration space is now $\ag$, the
space of suitably regular $U(1)$ connections modulo gauge
transformations.  Since the theory is Abelian, however, this space is
linear and can be identified with the space of magnetic fields
(satisfying the corresponding regularity conditions):
\begin{equation}
\ag = \lbrace \hbox{space of magnetic fields} \rbrace \ .
\end{equation}
Again, the gaussian measure $\mu$ (with covariance ${1\over
2}\Delta^{ {1 \over 2}}$) gives to the space of square
integrable cylindrical functions on $\ag$ the structure of a
pre-Hilbert space. As in the case of scalar fields, this space of
functions on the classical configuration space is not Cauchy complete.
Its Cauchy completion leads to states that can, in general, be
expressed only as functions on a space of distribution-valued magnetic
fields. Thus, again, the domain space of quantum states is larger than
the classical configuration space $\ag$.

In both cases, the domain space is the topological dual of the space
${\cal L}$ of smearing fields used in the definition of configuration
operators.  More precisely, the situation is as follows. Denote by
${\cal C}_{dist} \supset {\cal C}$ the space of (tempered)
distributions which is dual to the space of smearing fields ${\cal L}$
(endowed with the standard nuclear topology \cite{Ya}). Then, there is
a cylindrical measure $\tilde \mu$ on ${\cal C}_{dist}$ which is
equivalent to $\mu$ on the subspace of cylindrical functions but has,
contrary to $\mu$ on ${\cal C}$, an important technical property
($\sigma$-additivity; see Sect. 4) which ensures that the space
${L}^2({\cal C}_{dist},\tilde \mu)$ is complete. This is the Cauchy
completion of the pre-Hilbert space of square-integrable cylindrical
functions on ${\cal C}$. It is because of this that ${\cal C}_{dist}$,
rather than ${\cal C}$, is the domain space of quantum states.
Finally, the enlargement can not be ignored as ``an irrelevant
technicality'': although ${\cal C}$ is dense in ${\cal C}_{distr}$ in
the natural topology, the $\tilde\mu$-measure of ${\cal C}$ is in fact
zero! Thus, the measure is concentrated on distributional fields.
\footnote{In the case of Maxwell fields, for example, one can begin with
a function on $\cal C$. If it can not be continuously extended to
${\cal C}_{dist}$, however, it fails to define a (non-zero) quantum
state  in a natural way.}

\section{Integration on $\ag$}

Let us now consider non-Abelian gauge fields. Now, the classical
configuration space $\ag$ is genuinely non-linear. Nonetheless, the
``non-linear duality'' between (gauge equivalence classes of)
connections and loops will enable one, again, to introduce the notion
of cylindrical functions and cylindrical measures.

{} From now on, we will assume that the gauge group $G$ is $SU(N)$ or
$U(N)$ for some $N$ and let $\cal A$ be the (affine) space of
connections on a principal bundle $P(\Sigma, G)$ over a spatial
manifold $\Sigma$. (This $\Sigma$ is to be thought of as a Cauchy
slice in the Minkowskian signature and as the space-time itself in the
Euclidean signature.) It turns out that the final results are
independent of the initial choice of the bundle \cite{AL}. Let ${\cal
L}_{x_0}\Sigma$ be the space of continuous, piecewise analytic loops
on $\Sigma$ based at an arbitrarily chosen but fixed point $x_0$ on
$\Sigma$. Two loops, $\a , \b \in {\cal L}_{x_0}\Sigma$, are said to
be holonomically equivalent if, for every $A \in \cal A$, we have:
\begin{equation}
H(\a, A) = H (\b , A)   \ ,
\end{equation}
where $H(\a , A)$ denotes the holonomy of $A$ around $\a$ evaluated at
$x_0$; $H(\a , A) = {\cal P} exp(\oint_\a A)$.  The equivalence
classes of loops are called {\it hoops} (or $G$-hoops). For notational
simplicity, we will use lower case greek letters also to denote these
classes.  The set of all $G$-hoops forms a group called the {\it hoop
group} which is denoted by $\hg$. Traces of holonomies, being gauge
invariant, define for us the ``configuration observables'' $T_\a(A):=
\textstyle{1\over N} H(\a, A)$ on $\ag$ (where the trace is taken in
the fundamental representation of $SU(N)$ or $U(N)$). Thus, $\hg$ is
now the label set for configuration observables, the analog of the
Schwarz space ${\cal L}$ in the linear case.

Let us now introduce the notion of cylindrical functions on $\ag$.
This task appears difficult at first since $\ag$ does not have a
linear structure. However, we can carry it out essentially as in the
last section, using $\hg$ in place of ${\cal L}$
as the label set of observables (see \cite{AL,B93}).
We begin with the notion of independent hoops. Hoops $\{ \b_1 ,
\ldots
, \b_n \}$ will be said to be  independent if loop representatives of
loop equivalence classes can be chosen in such a way that each
contains an open segment that is traced exactly once and which
intersects any of the other representatives at most at a finite number
of points. Let now $S_n^*(\b_1 , \ldots , \b_n )$ denote the subgroup
of $\hg$ generated by a set of hoops,
$\{ \b_1 , \ldots , \b_n \}$.  As the notation
suggests these finitely generated subgroups will play the role of the
finite dimensional subspaces $V_n^*$ of Sect. 2.  Indeed, they define
{\it surjective} projections from $\ag$ to {\it finite dimensional
spaces} whose structure is well-understood \cite{AL} :
\begin{eqnarray}
\label{pron}
\pi_{\b_1 , \ldots , \b_n} ([A]) \ &   : & \ \ag \rightarrow G^n/Ad
\nonumber\\
\pi_{\b_1 , \ldots , \b_n} ([A]) & = & [H(\b_1, A), \ldots ,
 H(\b_n, A)],
\end{eqnarray}
where $G^n/Ad$ is the quotient of $G^n$ by the adjoint action of $G$.
Note the similarity with the construction presented in Sec. 2; the
role of quotient spaces $\R^n$ is now played by $G^n/Ad$. It is the
``universality'' of the structure of these quotients that lets us
repeat the constructions of Sect. 2.

Thus, cylindrical functions $f$ on $\ag$ can now be defined as the
functions that can be represented in the form
\begin{equation}
\label{cyag}
f = F \circ \pi_{\b_1 , \ldots , \b_n}   \ ,
\end{equation}
where $F$ is a function on $G^n/Ad$, \ $F: G^n/Ad \rightarrow \C$.  As
in Sect. 2 a function $f$, cylindrical with respect to $S^*_n({\b_1 ,
\ldots , \b_n})$ is cylindrical with respect to any group $\widetilde
S^*_m({\tilde \b_1 , \ldots , \tilde \b_m})$ that contains
$S^*_n({\b_1 , \ldots , \b_n})$. (In particular this is true for $m =
n$ in which case the two groups coincide but the the projection maps
$\pi_{\b_1 , \ldots , \b_n}$ and $\pi_{\tilde \b_1 , \ldots , \tilde
\b_n}$ are different if $\{ {\b_1 , \ldots , \b_n} \} \neq \{ {\tilde
\b_1 , \ldots , \tilde \b_n} \}$.)

A cylindrical measure $\mu$ on $\ag$ can now be defined, in a way that
is completely analogous to (\ref{cyme}), as a (consistent) family of
measures $\{ \mu_{\b_1 ,\ldots , \b_n} \}$ (one for each ordered set
of strongly independent hoops) on the compact finite dimensional spaces
$G^n/Ad$. Next, given a cylindrical function $ f = F \circ
\pi_{\b_1 , \ldots , \b_n}$ the integral of $f$ with respect to the
cylindrical measure $\mu$ on $\ag$ corresponding to the family $\{
\mu_{\b_1 , \ldots , \b_n}\}$ is now defined by
\begin{equation}
\label{cyma}
\int_{\ag} f([A]) d \mu([A]) = \int_{G^n/Ad} F([g_1, \ldots , g_n])
d \mu_{\b_1 , \ldots , \b_n}([g_1, \ldots , g_n])
\end{equation}
Finally, as before, we obtain the required consistency conditions on
the family $\{\mu_{\b_1 , \ldots , \b_n} \}$ of measures from this
definition: the family defines a cylindrical measure on
$\ag$ if and only if for every cylindrical function $f = F \circ
\pi_{\b_1 , \ldots , \b_n}$ such that $F$ is $\mu_{\b_1 , \ldots ,
\b_n}$-integrable and every group $\widehat S_m({\hat \b_1 , \ldots ,
\hat \b_n})$ containing $S_n({\b_1 , \ldots , \b_n})$, we have
\begin{equation}
\label{consg}
\int_{G^n/Ad} F([g_1, \ldots , g_n]) d\mu_{\b_1, \ldots ,
\b_n}([g_1, \ldots , g_n])    = \int_{G^m/Ad}
\widehat F([ g_1, \ldots , g_m])
d\mu_{\hat \b_1, \ldots , \hat \b_m}([g_1,
\ldots , g_m])   \ .
\end{equation}

The key question is if these consistency conditions can be met. The
answer is in the affirmative. A particularly natural solution $\mu^o$
to the consistency conditions (\ref{consg}) is the following
\cite{AL}: Let $\mu_n$ denote the measures on $G^n/Ad$ induced by the
normalized Haar measure in each of the factors $G$, and set
$\{\mu_{\b_1, \ldots, \b_n} \}$ to be:
\begin{equation}
\label{alm} \mu_{\b_1, \ldots , \b_n} = \mu_n
\end{equation}
for all sets $\{{\b_1, \ldots , \b_n} \}$.  Since this measure was
constructed without using any background structure on $\Sigma$ (such as
a metric), one would expects that it is diffeomorphism
($Diff(\Sigma)$-) invariant. This is indeed the case \cite{AL}.
Furthermore, an infinite set of $Diff(\Sigma)$-invariant solutions to
the conditions (\ref{consg}) has ben found \cite{B93}. These measures
can be thought of as generalized solutions to the diffeomorphism
constraint in the connection representation.

The measure $\mu^o$ has an additional important property of being
non-degenerate in the sense that for every continuous cylindrical
function $f \neq 0$, we have:
\begin{equation}
\label{fint}
\int_{\ag} \mid f([A]) \mid^2 d \mu^o([A]) > 0    \ .
\end{equation}
This concludes the discussion of the basic integration theory on $\ag$.
In the next section, we will present one perspective which provides an
intuitive grasp on this theory and especially on the properties of the
measure $\mu^o$.

\section{The measure and its support}

In principle, the measure $\mu^0$ on ${\cal A}/{\cal G}$ tells us
everything we need to know to construct the space of quantum states.
Thus, as in the Abelian case, we can begin with the space of
cylindrical, square integrable functions on ${\cal A}/{\cal G}$ as our
pre-Hilbert space and Cauchy complete it to obtain the full Hilbert
space of quantum states. One's task, then, is to represent this
completion as $L^2(\agb, \bar{\mu^o})$ for some $\agb$ that can be
thought of as a completion of $\ag$ and some measure
$\bar{\mu^o}$, which can be thought of an extension of $\mu^o$.
(While it is not a priori clear if $\agb$ is a genuine extension of
$\ag$ or equal to $\ag$ itself, we will see below that an extension is
indeed required.)  The space $\agb$ would be the analog of ${\cal
C}_{dist}$ from the Abelian case and its elements would be the
analogous of the ``distributional gauge-equivalent connections.''
These general expectations are indeed correct.

The detailed construction of $\agb$ and $\overline \mu^o$ can be carried out
in two different ways. In this section, we will sketch the method
developed in \cite{mm} which is based on projective limits. The
appendix will summarize the second method \cite{ai,AL}, based on the
representation theory of $C^\star$-algebras.

Many of the essential ideas for the construction of $\agb$ as a
projective limit have already been introduced in the last section.
Recall first that every $A$ in $\cal A$ defines an homomorphism from
the hoop group $\hg$ to the gauge group $G$
\begin{eqnarray}
h_A \ : \ \hg & \rightarrow & G \nonumber\\
\a & \mapsto & h_A(\a) = H(\a, A) \ .
\end{eqnarray}
Hence, every gauge equivalence class $[A]$ defines the following
equivalence class of homomorphisms:
  $$ h_{[A]} = [h_A] = [g^{-1} h_A g]\ , \quad \hbox{\rm for every}
  \ g \in G \ .  $$
Therefore, if $\cal F$ denotes the set of all finitely generated
subgroups of $\hg$, then every $S^*_n \in {\cal F}$ defines a
projection
\begin{eqnarray} \pi_{S_n^*} \ : \ \ag & \rightarrow \! &
{\cal Q}_{S_n^*} \equiv Hom(S_n^* , G)/Ad \nonumber\\ \! [A] & \mapsto
\! & [h_{A ; S^*_n}] \equiv [h_A \mid_{S^*_n}].
\end{eqnarray}
Thus, while two connections are gauge related if and only if the
holonomies they define around any hoop in $\hg$ are related by an
(hoop-independent) adjoint map, two connections have the same
projection under $\pi_{S^*_n}$ if and only if their holonomies are
related by an (hoop-independent) adjoint map for all hoops in $S^*_n$.
Since the projection map $\pi_{S^*_n}$ cares only about what the
connections do to the hoops in $S^*_n$, and since $S^*_n$ is generated
only by a finite number $(n)$ of independent hoops, it has a huge
kernel.  Indeed, it is not difficult to show that the spaces ${\cal
Q}_{S^*_n}$ are homeomorphic to finite-dimensional spaces $G^n/Ad$:
for every set $\{\b_1, \ldots , \b_n\} $ of independent hoops
generating $S^*_n$, the homeomorphism is given by
\begin{eqnarray}
f_{\b_1, \ldots , \b_n} \ : \ {\cal Q}_{S^*_n} & \rightarrow & G^n/Ad
\nonumber\\ \!  [h_{S^*_n}] & \mapsto & [h_{S^*_n}(\b_1), \ldots ,
h_{S^*_n}(\b_n)] \ .
\end{eqnarray}
Thus, the projections $\pi_{\b_1 , \ldots , \b_n}$ of Sect. 3 and the
$\pi_{S^*_n}$ defined above are related by
\begin{equation}
\pi_{\b_1, \ldots , \b_n} = f_{\b_1 , \ldots , \b_n }
\circ \pi_{S^*_n} \ ,
\end{equation}
where $\{\b_1 , \ldots , \b_n \}$ is any set of independent generators of
$S_n^*$.

These constructions imply that there is a relation between the
configuration space $\ag$ and ``consistent families" of (equivalence
classes of) homomorphisms $([h_{S^*}])$ where $h_{S^*} \in
Hom(S^*,G)$.  For, given an element of ${\cal A}/{\cal G}$, the
holonomies $h_{A;S^*}(\alpha) = H(\a, A)$ around every $\alpha \in
S^*$ for {\it any} $S^*$ define a homomorphism $[h_{A;S^*}]$ and the
family of homomorphisms obtained by varying $S^*\in {\cal F}$ is
consistent in the sense that, if $S^*_1 \subset S^*_2$, then
$[h_{A;S^*_1}]$ is simply the restriction to $S^*_1$ of $[h_{A;S^*_2}]$.
The key idea now is to consider the space $\cal Q$ of {\it all} such
consistent families $([h_{S^*}])_{S^* \in {\cal F}}$, whether or not
they arise from a connection.  This space is known as the ``projective
limit" of the spaces ${\cal Q}_{S^*} = Hom(S^*,G)/Ad$.  Thus, we have:
\begin{equation}
{\cal Q} = \{ ([h_{S^*}])_{S^* \in {\cal F}} : for \ S^*{}' \supset S^*,
[h_{S^*}] = [h_{S^*{}'}|_{S^*}] \equiv \pi_{S^*S^{*'}}([h_{S^{*'}}]) \}
\end{equation}

Note that there is close similarity between a projective limit space
and a direct product space since elements of each space can be
specified by giving an infinite number of components; in our case one
has to specify the entire family of (equivalence classes of)
homomorphisms $h_{S^*}$, for all $S^*\in {\cal F}$.  However, in the
case of a projective limit, these components describe the element in a
highly redundant way. In our example, all of the information in
$[h_{S^*}]$ is also contained in $[h_{S^*{}'}]$ for any $S^*{}'
\supset S^*$. This is why we must restrict ourselves only to {\it
consistent} families. One may picture the spaces ${\cal Q}_{S^*}$ as
forming the nodes of a kind of tree or net, in which each space ${\cal
Q}_{S^*_o}$ is viewed as being the root of a new tree that contains
all nodes ${\cal Q}_{S^*}$ for which $S^*$ is a subgroup of $S_o^*$.
Thus, while this tree has many final or smallest elements ---given by
${\cal Q}_{S^*}$ for those $S^*$ that are generated by a single
element--- the tree has no root or maximal element, since there is no
{\it finitely generated} subgroup of the hoop group that contains all
of the others.

If the space $\cal Q$ is itself thought of as a tree, then its
elements may be thought of as trees as well.  This time, however, the
nodes that make up the corresponding trees are not the entire spaces
${\cal Q}_{S^*}$, but single members of these spaces.  In other words,
since ${\cal Q}$ is a tree of sets, it may also be thought of as a set
of trees for which the $S^*$ node is some element taken from the set
(${\cal Q}_{S^*}$) that forms the $S^*$ node of ${\cal Q}$.  Just as
the nodes of the ${\cal Q}$-tree are linked together by the
projections $\pi_{S^*{}'S^*}$, so are the nodes of the trees $h \in
{\cal Q}$. This is because ${\cal Q}$ is exactly the set of all {\it
consistent} trees, satisfying $\pi_{S^*{}'S^*} ([h_{S^*}]) =
[h_{S^*{}'}]$ where $h = ([h_{S^*}])_{S^* \in {\cal F}}$.

Another bit of structure at our disposal is the ability to pick out,
from any tree $h\in {\cal Q}$, the node corresponding to any finitely
generated group $S^*$.  This is described mathematically by another
set of projection maps $\overline \pi_{S_0^*} : {\cal Q} \rightarrow
{\cal Q}_{S_0^*}$ which are defined so that $\overline \pi_{S_0^*}
([h_{S^*}]_{S^*}) = [h_{S_0^*}]$.

These considerations provide an intuitive picture of the projective
limit $\cal Q$. Let us now ask: How large is ${\cal Q}$?  We already
saw that each $[A]\in \ag$ defines a consistent family of
homomorphisms, and therefore an element of ${\cal Q}$. Furthermore,
distinct elements $[A_1]$ and $[A_2]$ of $\ag$ define distinct
families of homomorphisms since there is {\it some} hoop for which
$[h_{A_1}]$ is different from $[h_{A_2}]$. Thus $\ag$ is embedded in
${\cal Q}$. That ${\cal Q}$ is actually bigger is not so obvious and
the demonstration proceeds in three steps. First, one constructs
\cite{ai} a compact, Hausdorff space $\agb$ purely algebraically, as
the ``spectrum'' of the Abelian $C^\star$-algebra generated by the Wilson
loop functionals on $\ag$. One can show that $\ag$ is densely embedded
in $\agb$ (in the natural topology on $\agb$) \cite{AR}. In the second
step, one shows \cite{AL} that $\agb$ can also be obtained as the
space of all homomorphisms from the hoop group to the gauge group $G$
(modulo conjugation). Using this characterization of $\agb$, in the
third and the final step, one shows \cite{mm} that the projective
limit $\cal Q$ is in fact homeomorphic to $\agb$. Thus, $\ag$ is
densely embedded in $\cal Q$. From now on, we will simply identify
$\agb$ with $\cal Q$. Thus, the projective limit has provided us with a
completion of the configuration space $\ag$. Using the situation in
the linear case discussed in Sect. 2 as a guide, the limiting points
added to $\ag$ can be interpreted of as ``distributional gauge equivalent
connection'' \cite{AL}.

We can now define integration on $\agb$ in analogy with integration on
$\ag$ and use this integration theory to justify the use of $\agb$ as
the domain space of our quantum wavefunctions.

As before, let us introduce cylindrical functions $f$ of the form $f =
F \circ \overline \pi_{S^*}$ where $F$ is some
($\mu_{S^*}$-integrable) function on ${\cal Q}_{S^*}$.  We then define
the integral of $f$ as
\begin{equation}
\int_{\overline {{\cal A}/{\cal G}}} f \ d{\overline {\mu}}
= \int _{{\cal Q}^*} F \ d\mu_{S^*}   \ ,
\end{equation}
where $\{ \mu_{S^*} \}$ is a consistent family of measures, as in
Sect. 3. In particular, following \cite{AL}, one can consider the same
consistent family $\{ \mu_{S^*} \}$ as in (\ref{alm}). Let us denote
the corresponding cylindrical measure on $\agb$ as $\bar\mu^o$.

We will conclude this section by stating some properties of the
measure $\bar\mu^o$ which will provide an intuitive understanding of
the transition from $(\ag, \mu^o)$ to $(\agb, \bar\mu^o)$.  Recall
that a measure is an object that assigns a positive number (a sort of
``size") to sets that are well behaved (technically, ``measurable") in
a suitable sense.  The technical property that $\overline \mu^o$ possesses
but $\mu^o$ lacks is called $\sigma-additivity$.  This requires that,
given a countable collection $\{D_n\}$ of measurable non-intersecting
sets, $D_n \subset \agb$, the measure must satisfy:
\begin{equation}
\overline \mu^o(\cup_{n = 1}^\infty D_n) = \sum_{n=1}^\infty \overline
\mu^o(D_n).
\end{equation}
It is this property that guarantees that the the space of $\bar
\mu^o$-square integrable functions on $\agb$ is Cauchy-complete, i.e.,
is a Hilbert space, a property that is not enjoyed by the
corresponding functions on $\ag$. Indeed the fact that $\mu^0$ can not
be extended to a $\sigma$-additive measure on $\ag$ is equivalent to
the fact that there exist sequences of cylindrical sets
\footnote{A set $C$ in $\ag$ ($\agb$) is called cylindrical if it is
the inverse image $\pi_{S^*}^{-1}(B)$ ($\bar \pi_{S^*}^{-1}(B)$) of a
measurable set on ${\cal Q}_{S^*}$.}
$C_1 \subset C_2 \subset \ldots C_n \subset \ldots $ such that
even though $\cap_{n=1}^\infty C_n = \emptyset$, one has $\lim_{n
\rightarrow \infty} \mu^0(C_n) = q > 0$ (see Hopf theorem in,
e.g., \ \cite{Ya}).  By choosing the sets $C_n$ with adequate care (see
\cite{mm}) the corresponding sequences of characteristic functions
$\chi_{C_n}$ will be Cauchy but will not converge to a ($\mu^0$-square
integrable) function on $\ag$. Finally, viewed as functions on sets,
$\mu^o$ and $\overline \mu^o$ are related quite simply.  For any
cylindrical set $C \subset \agb$, we have:
\begin{equation}
\label{mu}
\mu^o(C \cap \ag) = \overline \mu^o(C)
\end{equation}
However, Eq. (\ref{mu}) holds {\it only} for cylindrical sets $C$ on
$\agb$.  In particular it does {\it not} hold for $C = \ag$ (notice that
$\ag$ is cylindrical in $\ag$ but not in $\agb$).  We have $\overline
\mu^o (\ag) = 0$ (see \cite{mm}) again in analogy with the linear case
and further justifying our use of $\agb$ as the domain space of quantum
states.

\section{Discussion}

The last four sections (and the Appendix) summarize some of the recent
mathematical developments on integration theory on $\agb$.  The
general strategy we followed has analogs in the integration theory on
infinite dimensional, linear spaces. However, all our constructions
were geared, from the very beginning, to the fact that the space $\ag$
is non-linear. We also outlined the construction of
$\overline \mu^0$, a non-degenerate,
diffeomorphism invariant measure on $\agb$.  The availability of such
measures was somewhat surprising at first because it was widely
believed that, just as translation-invariant measures do not exist in
infinite dimensional linear spaces, diffeomorphism invariant measures
would not exist on $\ag$. We also saw that while $\ag$ is the natural
configuration (or phase) space at the classical level for theories of
connections, the domain space for quantum states is its completion
$\agb$, which can be obtained as a projective limit.

The availability of this mathematical structure opens up avenues to
treat several problems of direct physical interest. First, the measure
$\bar\mu^o$ on $\agb$ can now be used to define the loop transform and
the loop representation rigorously. This is of particular interest to
quantum general relativity since $\bar\mu^o$ is diffeomorphism
invariant. However, the results presented here can not be used
directly in 4-dimensional, Lorentzian general relativity: although the
gauge group there is still $SU(2)$, the relevant connections are {\it
complex-valued} 1-forms which take values in the Lie algebra of
$SU(2)$. The projective limit techniques outlined here should,
however, go over to the more general case in a straightforward fashion
and enable us to construct $\agb$. Work is in progress to verify that
the integration theory goes through as well. Modulo this extension,
one can now establish the relation between quantum general relativity
and knot theory rigorously \cite{RS}: In the loop representation,
solutions to the diffeomorphism constraints of quantum general
relativity are functions of generalized knot classes of loops (where
the adjective ``generalized'' refers to the fact that loops are
allowed to have intersections, kinks and retracings). As mentioned in
the Introduction, the next step is to determine the additional
conditions which these functions should satisfy to be admissible as
candidate physical states. The required analysis involving rigged
Hilbert spaces (along the lines of \cite{Haj}) is, however, yet to be
attempted.  The availability of the loop transform should also enable
one to put the regularization of geometrical operators (such as the
area associated with a 2-surface) as well as ``weave states''
\cite{ars} on a rigorous footing. Finally, the techniques discussed
here together with those developed in connection with the differential
calculus on $\agb$
\cite{AL2} are likely to lead to a precise regularization of the
scalar (i.e. Hamiltonian) constraint of general relativity.

The present framework also provides a natural avenue to treat Yang-Mills
theories in the Hamiltonian as well as the Euclidean approach.
For definiteness, let us focus on Euclidean methods.  Since Yang-Mills
theories are {\it not} diffeomorphism invariant, the measure
$\bar\mu^o$ is not likely to be relevant in the final picture. One
must look for measures which are invariant only under the Euclidean
Poincar\'e group. A natural avenue is to use a suitable lattice
regularization. On a lattice, the framework constructed here can be
used directly. Furthermore, some simplifications arise. First of all,
since the lattice theories have only a finite number of degrees of
freedom, there is no distinction between $\ag$ and $\agb$. Next, since
the lattice action as well as all observables can be expressed in
terms of Wilson loop variables, the expectation values of observables
can be computed using the integration techniques for cylindrical
functions on $(\ag, \mu^o)$ associated with the lattice. The problem of
course is
that of the continuum limit. We believe that it should be possible to
extract some qualitative features of this limit using the general
integration theory developed here. More importantly, we believe that
these techniques may provide a natural extension of the standard
Osterwalder-Schrader axioms \cite{GJ,Riv}, in which the ``kinematical
non-linearities'' are taken into account from the beginning. The
resulting theory will not be (at least manifestly) local; the basic
variables will be Wilson loops rather than local variables such as
gauge fixed connections.  To obtain a specific continuum theory, we
need to choose an appropriate measure on $\agb$. There is, however, a
1-1 correspondence between measures on $\agb$ and functions on the
hoop group $\hg$ satisfying two rather simple algebraic conditions
\cite{ai,AL}.  This provides a new approach to the problem which may
well be fruitful.  Work is currently in progress in these directions
and some partial results are available in 2-dimensions \cite{AMMT}.

The framework does have a puzzling feature: It is based on the
assumption that the expectation values of Wilson-loop functionals
should be well-defined. This appears to be counter-intuitive because
the Wilson loop variables involve ``smearing'' only along one
dimension and, in free field theories in the Euclidean 4-space, one
normally finds that operators need to be smeared in all 4 dimensions.
It is clear therefore that the quantum theories we will find here will
not resemble free-field theories at all \cite{AI2}. Would they,
however, be viable? In 2-Euclidean dimensions, one can successfully
take the continuum limit of the lattice theory and show that, the
Wilson loop variables continue to have well-defined expectation values,
even though the loops are not smeared. Similarly, in the Hamiltonian
approach to 2+1 dimensional quantum general relativity, unsmeared
Wilson loop operators have a well-defined action on the Hilbert space
of physical states \cite{AAetal}. As one might expect, both these
treatments are
non-perturbative.  The result is quite striking especially in the
Yang-Mills context, where one is in the more familiar territory of
Poincar\'e invariant theories. However, in both these cases, the
theory has only a finite number of degrees of freedom and the nice
behavior of the Wilson loop variables may well be an artifact of this
feature. What about field theories with  genuinely infinite number of
degrees of freedom? Somewhat surprisingly, even in this case,
there is some evidence that the Wilson loop variables may be much
better behaved than what one might naively expect on the basis of free
field theories. It turns out that, in a perturbative treatment of
Yang-Mills theory in 4-Euclidean dimensions, after a multiplicative
renormalization, the Wilson loop variables again have finite
expectation values {\it to all orders} \cite{bns}. Again, no smearing
is involved. Results which hold to all orders in perturbation theory
are rare and generally reflect some intrinsic feature of
the underlying exact theory. ( As a side remark, it is also
interesting to note that, as in the framework presented in this
report, the natural arena for these perturbative calculations is again
provided by continuous, piecewise analytic loops.)  None of these
arguments are, of course, conclusive. But they suggest that it may
well be possible to construct a non-perturbative theory in 4-Euclidean
dimensions in which Wilson loop operators themselves are well-behaved.

There is an interesting, parallel development which may enable one to
 incorporate naturally ``smeared loops'' in the program outlined here
from the beginning. This stems from a recent extension
\cite{BGG} of the hoop group $\hg$ to an infinite dimensional Lie
group (of extended hoops) $\widetilde{\hg}$.  Mathematically, this
extension has a number of remarkable properties such as the existence
of a globally well defined exponential map establishing a one-to-one
correspondence between the group and its Lie algebra. Furthermore,
there exists a rigorous characterization \cite{JT} of
$\widetilde{\hg}$ as the spectrum of the (commutative non
co-commutative Hopf) algebra of certain functions on $\hg$ (the
``Chen-integrals''), which opens up new directions for exploring the
structure of $\widetilde{\hg}$ and using it in gauge theories.
Physically, these developments are significant because they allow us
to define a new class of gauge invariant functions of connections,
i.e., new algebras of functions on $\ag$. Thus, the results of
\cite{BGG,JT} provide powerful new tools and these have already found
significant applications to quantum gravity in 3+1-dimensions
\cite{BGP,GGP}. It is therefore natural to attempt to extend the
integration theory outlined here by replacing the hoop group $\hg$ by
$\widetilde{\hg}$. This extension, however, appears to be non-trivial:
some of the key results on the spectrum $\agb$ presented here depend
critically on geometric properties of loops and these are not shared
by extended loops. Nonetheless, if it should turn out that the
unsmeared Wilson loop variables fail to be well-defined in four
dimensional theories, an extension along these lines would be
indispensable.

We will conclude with a general remark. Some theoretical physicists,
especially the ones whose work is grounded more in geometry than
analysis, may come away with the impression that many of the points
discussed here are ``too technical'' to be directly relevant to
physics. After all, even in non-relativistic quantum mechanics, there
are many technical points --such as the precise domains of various
operators, the distinctions between self-adjoint and symmetric
operators, etc-- that are mathematically significant but can be
ignored in most physical applications. Are not the distinctions
between $\ag$ and $\agb$, between $\mu^o$ and $\bar\mu^o$ and between
$\hg$ and $\widetilde{\hg}$ similar? Do we really have to bother about
all the mathematical technicalities to extract physical predictions of
quantum general relativity or Yang-Mills theories?

The answer is that in the initial constructive stage, we {\it do} have
to worry about all these subtleties. Once a sufficient number of
results have been proven, we generally develop a good intuition and
can afford to gloss over subtleties and proceed to the level of care
that is appropriate for the problem. But what is appropriate is not
initially obvious. For example, in the early days of quantum
mechanics, working theoretical physicists found the distinctions
between finite and infinite dimensional spaces and between $n\times n$
matrices and operators such as $\hat{X}$ and $\hat{P}$ too technical
to make any ``real'' difference. However, they soon realized that the
difference is crucial for such basic issues as the existence of
operators $\hat{X}$ and $\hat{P}$ whose commutator is proportional to
identity. Now, we have absorbed such facts into our intuition. A
perhaps less obvious example comes from the Von-Neumann uniqueness
theorem which ensures that under certain mathematical restrictions,
every irreducible representation of the algebra generated by $U(a)
:=\exp ia\hat{X}$ and $V(b) := \exp ib\hat{P}$ is isomorphic with the
standard Schr\"odinger representation. There exist representations
which, for example, satisfy all the assumptions of the theorem except
for (weak) continuity of $U(a)$ in the parameter $a$ but which would
give us completely different physics. Theoretical physicists never
consider these, although they are distinguished from the Schr\"odinger
representation only by a subtle continuity condition. The delicate,
mathematical work was done by Von-Neumann and we have simply
incorporated the main message into our intuition. Based on this
intuition, without further worry, in every physical application of
non-relativistic quantum mechanics, we simply begin with the
Schr\"odinger equation. The situation, we believe, will be similar
with calculus on $\ag$. Once the theory is sufficiently developed, our
intuition will be sharpened, rules of thumb will be developed and we
will instinctively know which technical issues can be safely ignored
and which are critical for physical applications. Indeed, this is
occurring already to a certain extent.

{\sl Note Added:} After this work was completed and posted on an
electronic bulletin board, we learnt of another manifestly gauge
invariant approach to quantization, introduced already 13 years ago:
M. Asorey and P. K. Mitter, Commun.  Math. Phys. {\bf 80}, 43-58
(1981). Here one endows $\ag$ with the structure of a Hilbert manifold
and works on it directly, without carrying out an extension to $\agb$.
Its precise relation to our framework is not known.

\section{Appendix. Algebraic construction of the domain space of
quantum states}

In this appendix, we will provide an alternate construction of the
space $\agb$ using properties of Abelian $C^\star$-algebras. (For
details, see \cite{ai,AL}.)

The key ideas in this construction come from the Gel'fand
representation theory.  We will therefore begin with a brief
introduction to this theory in a simple context that often arises in
theoretical physics. Let $X$ be a compact Hausdorff space. The space
$C(X)$ of all complex-valued, continuous functions on $X$ obviously
forms a $\star$-algebra (where the $\star$-operation just corresponds
to complex-conjugation). We can equip it with the sup-norm:
   $$ ||f|| = \sup_{x\in X} |f(x)| \ .  $$
This structure equips $C(X)$ with the structure of an Abelian
$C^\star$-algebra with identity.  (Completion is not necessary since
$C(X)$ is already complete with respect to the sup-norm.) A
question one is often interested in is the following: Can one
reconstruct the topological space $X$ from the knowledge of the
$C^\star$-algebra $C(X)$ alone? The Gel'fand theory provides the
answer in the affirmative. The idea behind this ``reconstruction'' is
the following.  Each point $x\in X $ defines, trivially, a map from
$C(X)$ to the space of complex numbers: $x \ : \ f \mapsto f(x) \ ,
\ \forall f \in C(X)$. By inspection, this map is linear and respects the
multiplication and the $\star$ operation. Maps with these properties,
i.e.  \break
$\star$-homomorphisms from $C(X)$ to $\C$, are called simply
{\it complex homomorphisms}.  Thus, each point $x \in X$ provides us a
complex homomorphism from the $C^\star$ algebra $C(X)$ to the space of
complex numbers. It turns out that {\it every} (non zero) complex
homomorphism is of this type. Thus, the given topological space $X$
can be reconstructed algebraically as the space of complex
homomorphisms on $C(X)$. Furthermore, there is a powerful
generalization. For {\it any} Abelian $C^\star$ algebra $C$ with
identity, the space of all complex homomorphisms is a compact,
Hausdorff space (in a natural topology), called the {\it Gel'fand
spectrum of the algebra}, and denoted by $\Delta_C$. Finally, $C$ is
isomorphic to the $C^\star$-algebra of {\it all} continuous functions
on $\Delta_C$. Thus, every Abelian $C^\star$-algebra $C$ with identity
can be realized as the $C^\star$-algebra of all continuous functions
on a compact, Hausdorff space $\Delta_C$.

We can now turn to theories of connections. Let us begin with the
space $\cal A$ of all (suitably regular) connections and consider its
quotient $\ag$ by the local gauge group. Being gauge invariant, the
Wilson loop variables $T_\a$ (\ref{wlv}) can be regarded as functions
on $\ag$. For groups under consideration, they are bounded and
continuous in any of the standard topologies on $\ag$. Finally, they
separate the points of $\ag$: given any two elements of $\ag$, there
exists a hoop $\a$ such that $T_\a$ takes distinct values at these
points. Therefore, $T_\a$ can be thought of as ``generalized
coordinates.''  However, since $\ag$ is topologically non-trivial, the
Wilson variables are not independent; they are subject to the so-called
Mandelstam relations.

Nonetheless, the $T_\a$ naturally lead us to a $\star$-algebra, called
the holonomy algebra, and denoted by $\ha$ \cite{ai}. This is the
algebra of functions on $\ag$ consisting of finite linear combinations
(with complex coefficients) of finite products of $T_\a$'s, the
involution being given by complex conjugation. Let us introduce the
sup-norm
\begin{equation}
\parallel f \parallel = \sup_{[A] \in \ag}
\mid f([A]) \mid   \ , \quad  f \in {\cal HA}   .
\end{equation}
on $\ha$. Then, with these structures $\ha$ becomes a commutative
pre-$C^\star$-algebra with identity. The completion $\hab$, being a
sub-algebra of all continuous, bounded functions on $\ag$, is also an
algebra of functions on $\ag$. Therefore, every point $[A]$ in $\ag$
defines a non zero  complex homomorphism (or character) on the
algebra $\hab$ given by the evaluation on $[A]$, i.e., the
$\delta_{[A]}$-functional:
  $$\delta_{[A]} (f) = f([A]) \ .  $$
The set of all characters is the Gel'fand spectrum of the algebra. For
the $C^\star$-algebra $\hab$, this is the space $\agb$ we are seeking:
\begin{equation} \Delta_{\overline {\cal HA}} \equiv \agb
\end{equation}

As we saw above, if $X$ is a compact Hausdorff space and $C(X)$ the
$C^\star$-algebra of all continuous functions on $X$ then all the
characters of $C(X)$ are evaluation functionals and $X$ and
$\Delta_{C(X)}$ are homeomorphic. The homeomorphism is given by
$\d$ (see \cite{Ru})
\begin{eqnarray}
\d \ : \ X & \rightarrow & \Delta_{C(X)} \nonumber \\
x \in X \ , \ x & \mapsto& \d_x    \ .
\end{eqnarray}
However, since $\ag$ is not compact (and, moreover, since there is no
assurance that $\hab$ is the algebra of {\it all} bounded, continuous
functions in any of the standard topologies \cite{mv} on $\ag$) there
exist characters $[h]$ of $\hab$ for which there is {\it no}
equivalence class $[A] \in \ag$ such that $[h] =
\d_{[A]}$. (Some explicit examples of such more general characters are
given in \cite{AL}.) The spectrum $\agb$ is therefore larger than (the
$\d$-image of) $\ag$. In the natural Gel'fand topology however, as
usual, $\agb$ is a compact, Hausdorff space. Furthermore,
\begin{equation}
\d \ : \ \ag \rightarrow \agb
\end{equation}
is a dense embedding (see \cite{AR}). This might be surprising at
first since in any of the standard topologies \cite{mv}, $\ag$ is not
even locally compact; the Gel'fand topology on $\agb$ is quite
different from any of these.

The starting point in this construction is the Wilson loop variables.
As far as the holonomy algebra or its spectrum are concerned, we could
have let the loops be, say, continuous and piecewise $C^1$. However,
at this stage of the development, (for non-Abelian gauge groups) all
the subsequent constructions require a stronger assumption: the loops
have to be continuous and piecewise {\it analytic}. This is necessary
because a key step in the subsequent   integration
theory involves the decomposition of any given loop into a finite
number of {\it independent} loops, where a set of loops is regarded as
independent if each loop in the set has a segment which is traced
exactly once and which is not shared by any other loop. For piecewise
analytic loops, it is straightforward to achieve such a decomposition,
while for piecewise smooth ones, it is not clear that a decomposition
always exists.  In the piecewise analytic category, one can give
another characterization of the spectrum $\agb$ which is often more
convenient to use in practice \cite{AL}: there is a one to one
correspondence between elements $[h]$ of $\agb$ and (equivalence
classes of) homomorphisms $[H]$ from the hoop group $\hg$ to the gauge
group $G$ (where two homomorphisms $H_1$ and $H_2$ are regarded as
being equivalent if $H_1(\gamma) = g^{-1}\cdot H_2(\gamma)\cdot g$ for
some hoop independent element $g$ of the gauge group $G$). It is this
characterization that most easily relates the spectrum $\agb$ to the
projective limit discussed in the main text.

Finally, since $\agb$ is the spectrum of $\hab$, it follows from the
general theory that $\hab$ is isomorphic to the algebra $C(\agb)$ of
all continuous functions on $\agb$. The isomorphism is given by the
``Gel'fand transform'':
\begin{eqnarray}
\tilde {} \ : \ \overline {\cal HA} & \rightarrow & C(\agb)
\nonumber\\
 f & \mapsto & \tilde{f}  , \\
\tilde{f}([h]) &=& [h]({f}) \ , \quad for \ [h] \in \agb \ .
\nonumber
\end{eqnarray}
Since $\agb$ is a compact, Hausdorff space, a major simplification
arises in quantum theory: we can directly use the Riesz representation
theorem which establishes a one-to-one correspondence between positive
linear functionals on $\hab$ and regular Borel measures on $\agb$ (see
\cite{R2}). Furthermore, for every regular Borel measure $\hat
\mu$ on $\agb$ there is \cite{ai} a cyclic, $\star$-representation of
$\hab$ by (bounded) multiplication operators on the Hilbert space
$L^2(\agb, \overline \mu)$.  Conversely, every cyclic representation
is unitarily equivalent to a representation of this type.  Finally,
there is a one-to-one correspondence between consistent families of
measures $\{ \mu_{S^*} \}$ and regular Borel measures $\widehat \mu$
on $\agb$: The measure $\widehat \mu$ is an extension of the (unique)
cylindrical measure $\overline \mu$ corresponding to the family $\{
\mu_{S^*} \}$.

\vskip 1cm

\noindent {\bf Acknowledgements}
\vskip .5cm
The authors would like to express their appreciation to John Baez,
Chris Isham, Jerzy Lewandowski and Thomas Thiemann for their long
history of useful comments and discussions.


\begin{thebibliography}{99}


\bibitem{GJ} J. Glimm and A. Jaffe, ``Quantum physics", Springer-Verlag,
New York, 1987

\bibitem{Riv} V. Rivasseau, ``From perturbative to constructive
renormalization", Princeton University Press, Princeton, 1991

\bibitem{gi} E. S. Abers and B.W. Lee, Phys. Rep. {\bf 9} (1973) 1

\bibitem{JZ} J. Zinn-Justin, ``Quantum field theory and critical
phenomena", Clarendon Press, Oxford, 1993

\bibitem{QCD}
M. Creutz, ``Quarks, gluons and lattices", Cambridge Univ. Press,
London, 1983

E.V. Shuryak, ``The QCD vacuum, hadrons
and the superdense matter", World Scientific, Singapore, 1988

\bibitem{AA87} A. Ashtekar, Phys. Rev. {\bf D36} (1987) 1587-1602

\bibitem{RS} C. Rovelli and L. Smolin, Nucl. Phys. {\bf B331} (1990)
 80-152

\bibitem{BGP} B. Br\"ugmann, R. Gambini and J. Pullin,
Phys. Rev. Lett. {\bf 68} (1992) 431-434

\bibitem{W} E. Witten, Comm. Math. Phys. {\bf 121} (1989) 351-399

\bibitem{CS} Ch. Nash, ``Differential topology and
 quantum field theory", Academic Press, London, 1991

\bibitem{AAetal} A. Ashtekar, V. Husain, C. Rovelli, J. Samuel and
L. Smolin, Class. Quan. Grav. {\bf 6} (1989) L185-L193

\bibitem{YM0} S. Chung, M. Fukuma and A. Shapere, ``Structure
of topological lattice field theories in three dimensions",
Preprint CLNS 93/1200, hep-th/9305080

\bibitem{Baez1}
 J. Baez, ``Strings, loops, knots and gauge fields",
to appear in {\it Knots
and quantum gravity}, J. Baez (ed), Oxford University Press

\bibitem{AR2}
A. Ashtekar and J.D. Romano, Phys. Lett. {\bf B229} (1989) 56-60

\bibitem{Haj} P. Haji$\check{\rm c}$ek, ``Quantization
of systems with constraints", to appear in {\it The canonical
formalism in classical and quantum gravity}, J. Ehlers and
H. Friedrich (eds), Springer-Verlag

\bibitem{GT} R. Gambini and A. Trias,   Phys. Rev. {\bf D22} (1980)
1380-1384

\bibitem{GT2} R. Gambini and A. Trias,  Nucl. Phys. {\bf B278} (1986)
436-448

\bibitem{B} B. Br\"ugmann,  Phys. Rev. {\bf D43} (1991) 566-579

\bibitem{Lo} R. Loll, Nucl. Phys. {\bf B368} (1992) 121-142

\bibitem{ai}
A.~Ashtekar and C.J. Isham, Class. Quan. Grav. {\bf 9} (1992) 1069-1100

\bibitem{AL}
A. Ashtekar and J. Lewandowski, ``Representation theory of
analytic holonomy $C^\star$ algebras", to appear in {\it Knots
and quantum gravity}, J. Baez (ed), Oxford University Press

\bibitem{B93} J. Baez, ``Diffeomorphism-invariant generalized
measures on the space of connections modulo gauge transformations",
hep-th/9305045, to appear in the Proceedings of
the conference on quantum topology, L. Crane and D. Yetter (eds).

\bibitem{mm} D.~Marolf and J. M. Mour\~ao,
``On the support of the Ashtekar-Lewandowski measure"
submitted to {\it Commun. Math. Phys.}

\bibitem{AL2} A. Ashtekar and J. Lewandowski, ``Differential
calculus in the space of connections modulo gauge transformations",
Preprint CGPG

\bibitem{GV} I. Gel'fand and N. Vilenkin, ``Generalized functions",
Vol. IV, Academic Press, New York, 1964

\bibitem{Ya} Y. Yamasaki, ``Measures on infinite dimensional spaces",
World Scientific, Singapore, 1985

\bibitem{AR} A. Rendall, Class. Quant. Grav. {\bf 10} (1993) 605-608

\bibitem{ars} A. Ashtekar, C. Rovelli and L. Smolin, Phys. Rev. Lett.
{\bf 69} (1992) 237-240

\bibitem{AMMT} A. Ashtekar, D. Marolf, J.M. Mour\~{a}o and
T. Thiemann, work in progress

\bibitem{AI2} A. Ashtekar and C.J. Isham, Phys. Lett. {\bf B274}
(1992) 393-398

\bibitem{bns} R. A. Brandt, F. Neri and M. Sato, Phys. Rev. {\bf D24}
(1981) 879-902

\bibitem{BGG} Di Bartolo, R. Gambini and J. Griego, Comm. Math. Phys.
{\bf 158} (1993) 217-240

\bibitem{JT} J.N. Tavares, ``Chen integrals, generalized loops
and loop calculus", Lisbon preprint IST 93, to appear in Int. J. Mod.
Phys. {\bf A}

\bibitem{GGP} R. Gambini, Griego, J. Pullin, ``Extended loops:
a new arena for nonperturbative quantum gravity", Preprint
CGPG-93/12-1, gr-qc/9312029

\bibitem{Ru} W. Rudin, ``Functional analysis", McGraw-Hill Book
Company, New York, 1973

\bibitem{mv} P.K. Mitter and C. M. Viallet, Commun. Math. Phys.
{\bf 79} 43-58

\bibitem{R2} W. Rudin, ``Real and complex analysis", McGraw-Hill Book
Company, New York, 1987


\end{thebibliography}
\end{document}